\documentclass[aps,prl,showpacs, twocolumn, superscriptaddress]{revtex4}
\usepackage[utf8]{inputenc}
\usepackage[english]{babel}
\usepackage{graphicx}
\usepackage{color}
\usepackage{bm}
\usepackage{amssymb}
\usepackage[intlimits]{amsmath}
\usepackage{amsmath}
\usepackage{subfigure}
\usepackage{cancel}
\newcommand{\be}{\begin{equation}}
\newcommand{\ee}{\end{equation}}
\newcommand{\ben}{\begin{equation*}}
\newcommand{\een}{\end{equation*}}

\begin{document}

\title{
Bound states of skyrmions and merons near the Lifshitz point
}
\author{Y. A. Kharkov}
\affiliation{School of Physics, University of New South Wales, Sydney 2052,
Australia}
\author{O.P. Sushkov}
\affiliation{School of Physics, University of New South Wales, Sydney 2052,
Australia}
\author{M. Mostovoy}
\affiliation{Zernike Institute for Advanced Materials, University of Groningen, Nijenborgh, Groningen, Netherlands.}

\begin{abstract}
We study topological defects in anisotropic ferromagnets with competing interactions near the Lifshitz point. We show that skyrmions and bi-merons are stable in a large part of the phase diagram. We calculate skyrmion-skyrmion and meron-meron interactions and show that skyrmions
attract each other and form ring-shaped bound states in a zero magnetic field. At the Lifshitz point merons carrying a fractional topological charge become deconfined. These results imply that unusual topological excitations may exist in weakly frustrated magnets with conventional crystal lattices.
\end{abstract}
\maketitle
{\em Introduction: }
%
Some fifty years ago Tony Skyrme identified topologically stable ``hedgehog''-like configurations of the meson field with baryons, such as proton and neutron~\cite{Skyrme1962}.
The ensuing theoretical work showed that skyrmions indeed provide a semiquantitative description of physical properties of baryons and their interactions~\cite{Zahed1986}.
Multi-skyrmion bound states describe ground states and low-energy excitations of atomic
nuclei~\cite{Battye2016}. Periodic crystals of skyrmions and half-skyrmions were used to model nuclear matter~\cite{Klebanov1985,Goldhaber1987,Kugler1989}.

Two-dimensional analogues of Skyrme's skyrmions are relevant topological excitations in many condensed matter systems~\cite{Nagaosa2013}, such as Quantum Hall magnets~\cite{Sondhi1993,Barett1995}, spinor Bose-Einstein condensates~\cite{Ho1998}, chiral liquid crystals~\cite{Fukuda2011}, and chiral magnets~\cite{Bogdanov1989}, which provide the playground for experimental studies of skyrmions. 
Skyrmion crystals and isolated skyrmions in chiral magnets can be observed by neutron scattering and Lorentz microscopy~\cite{Muhlbauer2009, Yu2010}, and controlled by ultralow electric currents~\cite{Jonietz2010, Yu2012}, applied electric fields~\cite{White2014,Seki2012}, and thermal gradients  \cite{Mochizuki2014}, which opened a new active field of research  on skyrmion-based magnetic memories~\cite{Fert2013, Iwasaki2013, Zhou2014, Jiang2015, Moreau2016, Woo2016}.
Half-skyrmions (or merons) carying half-integer topological charge  were also discussed theoretically in the context of quantum Hall systems~\cite{Brey1996}, bilayer graphene~\cite{Cote2010} and chiral magnets~\cite{Ezawa2011, Lin2015}, but so far they eluded experimental detection.

Here, we are interested in magnetic multi-skyrmion and multi-meron  configurations with a large topological charge, $Q$. An example is the skyrmion crystal in chiral magnets. However, skyrmions in the crystal can hardly be considered as independent particle-like objects, since to a good approximation this state is a superposition of three spin spirals plus a uniform magnetization~\cite{Muhlbauer2009}. Isolated skyrmions appear under an applied magnetic field that suppresses modulated spiral and skyrmion crystal phases and induces a collinear ferromagnetic (FM) state. Skyrmions in chiral magnets repel each other~\cite{Roessler2011}, so that multi-skyrmion states are merely a gas of elementary skyrmions with $Q = \pm 1$.

 It was recently suggested that skyrmion crystals and isolated skyrmions can also exist in frustrated magnets with conventional centrosymmetric lattices, where they are stabilized by competing ferromagnetic and antiferromagnetic (AFM) exchange interactions~\cite{Okubo2012, Leonov2015, Lin2016}. In frustrated magnets, the skyrmion-skyrmion interaction potential changes sign as a function of the distance between skyrmions, which makes possible  formation of skyrmion clusters as well as  rotationally symmetric skyrmions with the topological charge $Q = \pm 2$~\cite{Leonov2015}.
 
We note that topological excitations in frustrated magnets can be stable even in zero magnetic field. Consider a magnet in which the  degree of frustration described by the parameter $f$ that can be varied, e.g.  by an applied pressure or a chemical substitution. The phase diagram of such  magnets often contains the so-called Lifshitz point (LP), $f = f_\ast$, which separates the uniform FM state ($f < f_\ast$) from periodically modulated phases ($f > f_\ast$). The behavior close to the LP was recently discussed in the context of Bose condensation of multi-magnon bound states in quantum low-dimensional systems~\cite{Balents2016}.  Skyrmions, which can be considered as bound states of a large number of magnons, so far were studied in the strongly frustrated regime $(f > f_\ast)$. In this Letter, we focus on the ``underfrustrated'' side of the LP and show that skyrmions are stable in a large interval of $f < f_{\ast}$. We show that elementary skyrmions attract each other and can form bound states with an arbitrarily large $Q$. Surprisingly, despite the attraction, skyrmions do not aggregate into clusters. Instead, they form topological ring-shaped domain walls.  
 
 The aforementioned unusual multi-$Q$ states appear in easy-axis magnets. An easy-plane anisotropy forces skyrmion to transform into a bound pair of merons with opposite vorticities, each carrying topological charge $Q = \frac{1}{2}$.  The lowest-energy multi-meron configuration is a square lattice of merons with alternating vorticities. Our results show that stability of skyrmions and merons does not require strong magnetic frustration, implying that these exotic topological excitations with interesting physical properties can exist in already known magnetic materials. In addition, we find a number of striking similarities between multi-Q skyrmions in condensed matter and nuclear physics.


{\em The Model: }
We consider classical spins on a square lattice with competing exchange interactions and magnetic anisotropy. The energy of the model is 
\begin{eqnarray}\label{eq:J-K_Hamilt}
E= &\!\!-\!\!&J_1\sum_{\langle i,j\rangle} {\bf S}_i\cdot {\bf S}_j + J_2 \!\sum_{\langle\langle i,j\rangle\rangle}\! {\bf S}_i\cdot {\bf S}_j + J_3\!\! \sum_{\langle\langle\langle i,j\rangle\rangle\rangle}\!\! {\bf S}_i\cdot {\bf S}_j \nonumber\\ &\!\!+\!\!& \frac{K}{2} \sum_i \left(1-\left(S^z_i\right)^2\right),
\end{eqnarray}
where $\bm S_i$ is the spin of unit length at the lattice site $i$ and the first, second and third terms describe, respectively, FM nearest-neighbor and  AFM second- and third-neighbor exchange interactions   ($J_1,J_2,J_3>0$). The $z$ axis is normal to the lattice plane (the $xy$ plane) and $K$ is the strength of the single-ion magnetic anisotropy of easy axis ($K>0$) or easy plane  $(K<0)$ type.
In what follows, energy is measured in units of $J_1 = 1$ and  distances are measured in units of the lattice constant.

\begin{figure}[h!]
\includegraphics[scale=0.20]{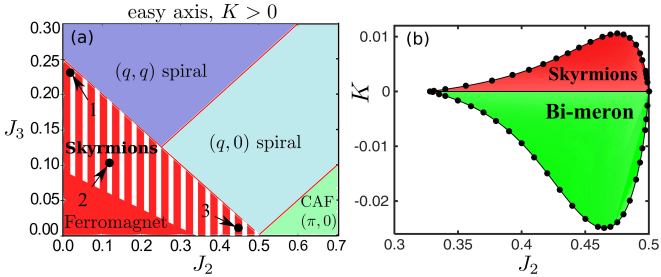}
\caption{(a) $J_2-J_3$ phase diagram of the square lattice frustrated magnet with a weak  magnetic anisotropy. Stripes show the stability region of the $Q = 1$ skyrmion in the FM phase, for $K = 10^{-3}$. (b) $J_2 - K$ stability diagram of skyrmions and bi-merons with $Q=1$, for $J_3 = 0$.
}\label{fig:phase_diagr}
\end{figure}



For slowly varying spin textures, Eq.(\ref{eq:J-K_Hamilt}) is equivalent  to the coninuum model (see e.g. Ref.~[\onlinecite{Milstein2011}]),
\begin{eqnarray}
E &=&\frac{1}{2}\int d^2 r \left[ \rho (\partial_i \bm S)^2 + b_1 \left( (\partial_x^2 \bm S)^2 + (\partial_y^2 \bm S)^2\right)
\right.\nonumber\\
 &+& \left. b_2 \partial^2_x \bm S \cdot \partial^2_y \bm S  + K(1-S_z^2) \right], 
 \label{eq:Energy[n]}
\end{eqnarray}
where $\rho= J_1-2J_2-4J_3$, $b_1=\frac{1}{12}(-J_1+2 J_2+16 J_3)$, $b_2=J_2$ and $i=x,y$. The first term in Eq.~(\ref{eq:Energy[n]}) is the $O(3)$ nonlinear sigma model of an isotropic two-dimensional ferromagnet. The spin stiffness, $\rho$, plays the role of $f_\ast - f$: in the FM state $\rho > 0$, at the Lifshitz point $\rho$ vanishes and for $\rho < 0$, the system has  either a spiral or a columnar antiferromagnetic (CAF) ground state~\cite{Seabra2015}, as shown in Fig. \ref{fig:phase_diagr}(a) \cite{footnote1}.  The fourth-order terms in gradients of $\bm S$ stabilize the spiral state and  determine its  wave vector  provided that $b_1 > 0$ and $b_1 + b_2/2 > 0$.  These terms also stabilize skyrmions and merons in the FM state.



{\em Skyrmions: } The  nonlinear sigma model with $\rho > 0$ allows for analytic expression for skyrmions with an arbitrary $Q$  found by Belavin and Polyakov~\cite{Belavin1975}. In the conformally  invariant sigma model skyrmions have no internal length scale: the energy,  $E_{Q}$, of the skyrmion with topological charge $Q$  is  $4\pi \rho |Q|$ independent of the skyrmion size.  

The radius, $R$, of the skyrmion with $Q = \pm 1$ (the elementary skyrmion) in frustrated magnets is determined by the competition between the fourth-order terms favoring infinite $R$ and the easy-axis  anisotropy that tends to shrink the skyrmion. The dimensional analysis shows that $R \sim \left[ \frac{b}{K}\right]^{1/4}$, where $b$ is a linear combination of $b_1$ and $b_2$ \cite{footnote2}. Skyrmion stability requires $\rho > 0$ and $b>0$. In particular, skyrmions in the Heisenberg model with the nearest-neighbor interactions only are unstable. Figure~\ref{fig:phase_diagr}(b) shows the  stability region of the elementary skyrmion in the $J_2-K$ plane calculated numerically for $J_3 = 0$ \cite{footnote2}. Note that skyrmions are stable quite far from the LP $J_2 = 1/2$.

\begin{figure}[h!]
\includegraphics[scale=0.29]{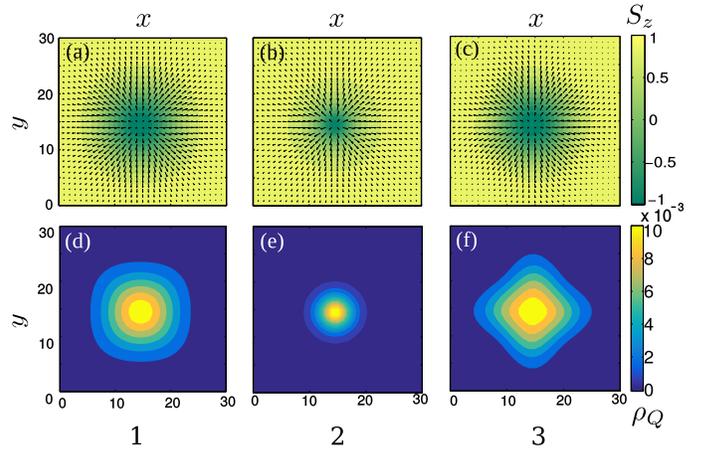}
\caption{(a-c) Elementary skyrmion at the points $1,2,3$ on the phase diagram Fig. \ref{fig:phase_diagr}a. Arrows show in-plane spin components, color indicates $S_z$. (d-f) The corresponding contour plots of the topological density,  $\rho_Q(x,y)$.}\label{fig:skyrmion_shapes}
\end{figure}

The skyrmion shape is controlled by the parameters $b_1$ and $b_2$, as shown in Figs.~\ref{fig:skyrmion_shapes}(a,b,c) and the corresponding contour plots of the topological  charge density,  
$
\rho_Q(x,y) = \frac{1}{4\pi} \bm S \cdot [\partial_x \bm S \times \partial_y \bm S] \label{eq:rho_Q}
$ 
(Figs.~\ref{fig:skyrmion_shapes}(d,e,f)). 
The square-shaped skyrmions are observed close to the LPs  $(J_2,J_3)=(1/2,0)$ and $(J_2,J_3)=(0,1/4)$. For  $J_2 > 2J_3$, the FM phase  transforms into the spiral state with the wave vector, $\bm q$, parallel to the square lattice axes, in which case the skyrmion has the shape shown in  Figs.~\ref{fig:skyrmion_shapes}(a,d). For $J_2 < 2J_3$,  $\bm q$ is along the diagonals of squares and the skyrmion has the shape shown in  Figs.~\ref{fig:skyrmion_shapes}(c,f).

\begin{figure}[h]
\includegraphics[scale=0.3]{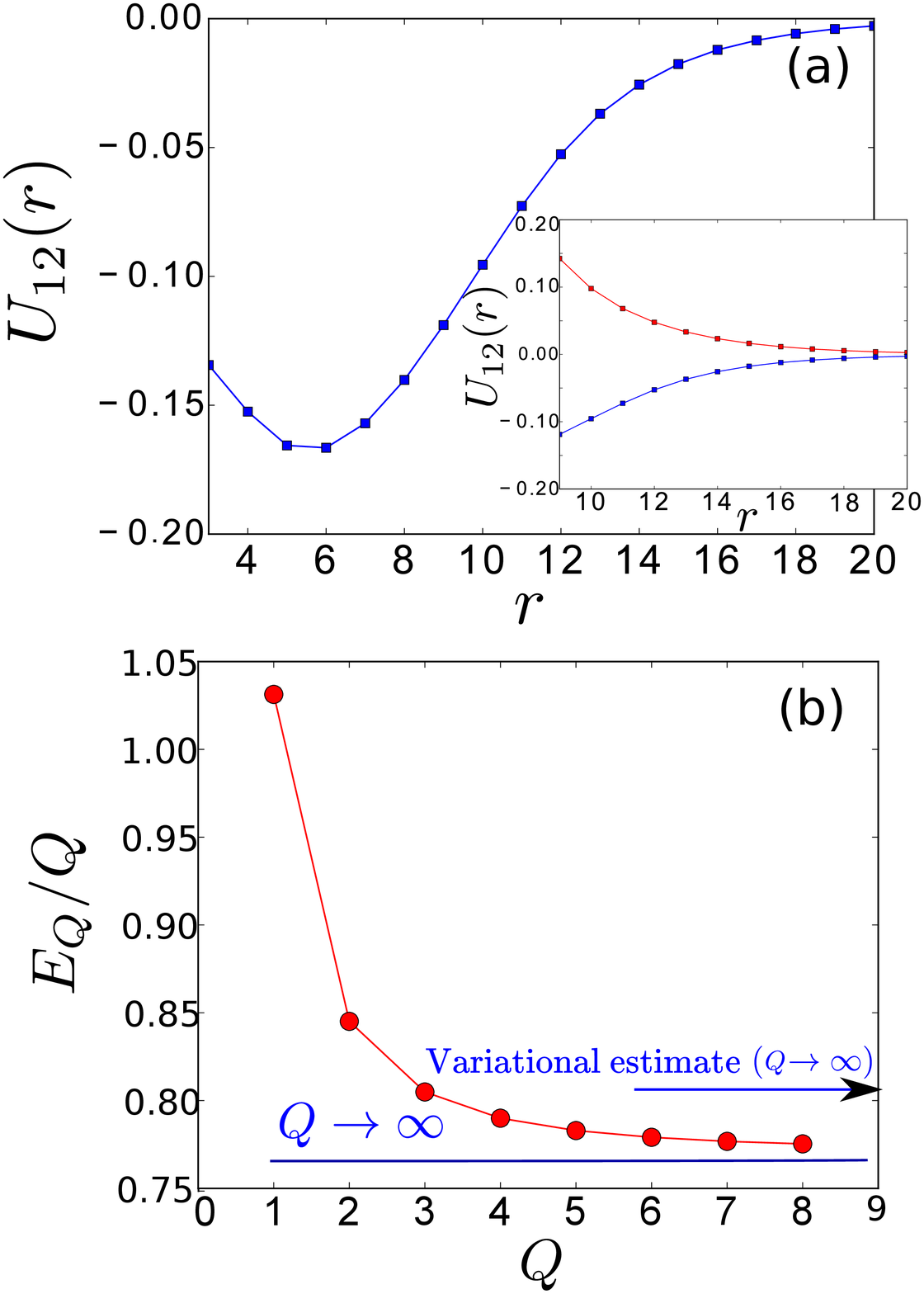}
\caption{  
(a) Potential energy, $U_{12}(r)$, of interaction between two skyrmions as a function of distance between the centers of the skyrmions for equal helicities, $\chi_1=\chi_2$, (blue line) and for opposite helicities, $\chi_1-\chi_2 = \pi$ (red line).
(b) Energy per skyrmion,  $E_Q/Q$, in the skyrmion ring. The calculations were preformed for $J_2=0.2$, $J_3=0.149$, and $K=0.01$.
 }\label{fig:Q=123}
\end{figure}

An important difference between skyrmions for positive and negative spin stiffness is the form of the skyrmion-skyrmion interaction potential, $U_{12}(r)$. For $\rho < 0$, the potential oscillates, which leads to repulsion or attraction depending on the distance,  $r$, between skyrmions~\cite{Leonov2015, Lin2016}. 
Similar considerations show that for $\rho > 0$, $U_{12}(r)$ remains positive and decreases monotonically at large $r$. This is also the case for skyrmions in chiral magnets which repel each other~\cite{Roessler2011,Rozsa2016} because they all have the same helicity  angle describing the direction of the in-plane spin components~\cite{Nagaosa2013}. In easy-axis magnets with competing interactions, the skyrmion helicity is arbitrary and the repulsion for equal helicities changes to attraction for opposite helicities  (see Fig.~\ref{fig:Q=123}(a)). 

\begin{figure}[h!]
\includegraphics[scale=0.44]{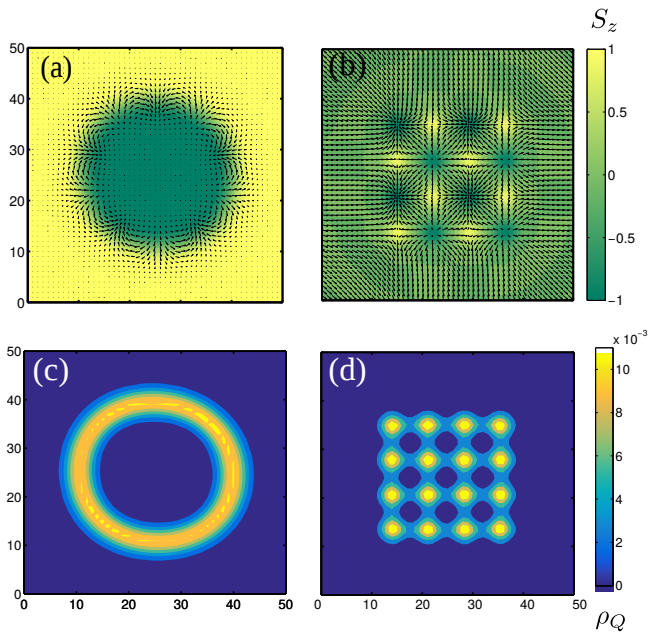}
\caption{(a)
The spin configuration of the skyrmion ring with $Q=-6$ in the frustrated magnet with an easy axis anisotropy $K=0.01$ and (c) the corresponding topological charge density distribution, $\rho_Q(x,y)$. 
(b) A meron cluster with a square lattice of vortices and antivortices minimizing energy for topological charge $Q=-8$ and the easy plane anisotropy $K=-0.01$; (d) the corresponsing $\rho_Q(x,y)$. Other parameters of these simulations are: $J_2=0.2$ and $\rho = 3 \cdot 10^{-3}$.
}\label{fig:multi-Q}
\end{figure}

Because of the attraction, a multi-$Q$ skyrmion has a  lower energy than $Q$ elementary skyrmions and can be considered as their bound state. The fact that $U_{12}(r)$ has minimum at $r = r_0$ (see Fig.~\ref{fig:Q=123}(a)) suggests that the skyrmion with a large $Q$ occupies the  area $\sim Q \pi r_0^2$, so that the skyrmion radius $R \sim r_0 Q^{1/2}$.  Surprisingly, this is not the case: the topological charge and energy densities of the multi-$Q$ skyrmion are concentrated in a ring of radius $R \sim r_0 Q$ (see Figs.~\ref{fig:multi-Q}(a,c)).  

Figure~\ref{fig:Q=123}b shows that the energy per skyrmion, $E_Q/Q$, decreases with increasing $Q$ and approaches a constant, because the width of the ring and the length of the ring segment occupied by one skyrmion become $Q$-independent. The energy of skyrmion in the ring is significantly lower than that of the elementary skyrmion. This ``mass defect'' drives the fusion of skyrmions, which increases the magnitude of the skyrmion magnetic moment, $M_z = \sum_i (S_i^z -1) < 0$, counted from the positive magnetic moment of the FM state: for $Q$ elementary skyrmions $M_z \propto - Q$,  whereas  for the ring with topological charge $Q$, $M_z \propto - Q^2$. A  magnetic field applied in the positive $z$ direction would lead to fission of multi-$Q$ skyrmions into skyrmions with smaller topological charges.

For  $Q \gg 1$, we can neglect the ring curvature and consider a straight domain wall with the spiral spin structure, $\bm S = (\sin \theta(x) \cos qy, \sin \theta(x) \sin qy, \cos \theta (x))$, separating the $S_z=-1$ FM state at $x<0$ from the $S_z=+1$ FM state at $x>0$. The length of the wall in the $y$ direction is $L_y = 2\pi R = \frac{2\pi Q}{q}$. Using a variational  Ansatz for the wall shape,  $\cos \theta(x) = - \tanh (\varkappa x)$, where $\varkappa$ is the inverse domain wall width, we obtain 
\begin{equation}\label{eq:dwenergy2}
\frac{E_Q}{Q} = \frac{2\pi}{q\varkappa} 
\left[
\rho \left(q^2+\varkappa^2 \right) + 
b_1 \left(q^4+\varkappa^4\right)
 + 
 \frac{b_2}{3} q^2 \varkappa^2
 + K  \right].
\end{equation}
Minimization with respect to $q$ and $\kappa$ gives 
$
\kappa = |q| = \left[ \frac{K}{2b_1+\frac{1}{3}b_2} \right]^{\frac{1}{4}}
$
and
\begin{equation}\label{eq:E_Q}
\frac{E_{Q}}{|Q|} = 4 \pi \left(\rho + \sqrt{K\left(2b_1+\frac{1}{3}b_2\right)} \right).
\end{equation}
Note that the first term in Eq.(\ref{eq:E_Q}) is the lower bound for the  energy of the multi-$Q$ skyrmion in the nonlinear sigma model~\cite{Belavin1975} and that $\varkappa$, $q$ and $E_{Q}/|Q|$ are indeed independent of $Q$. The domain wall stability requires $6b_1+b_2 > 0$ (or $4 J_2 + 16 J_3 > J_1$) and this result can be shown to be independent of the orientation of the wall with respect to the crystal axes. The binding energy makes multi-$Q$ skyrmions more stable than elementary skyrmions \cite{footnote2}.

\begin{figure}[h]
\includegraphics[scale=0.285]{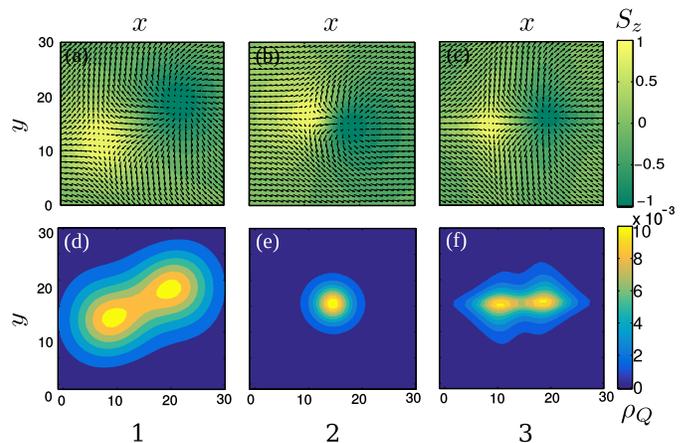}
\caption{(a-c) Deconfinement of meron pairs in a frustrated magnet with an easy plane anisotropy $K<0$ for $J_2$ and $J_3$ at the points $(1,2,3)$ on the phase diagram in Fig. \ref{fig:phase_diagr}a. 
(d-f) Topological density $\rho_Q$ for the spin configurations in (a-c). 
In panels (a,d) and (c,f) the system is close to the FM-spiral phase boundary, which results in the topological ``fractionalization'', i.e. a spatial separation of merons.
}\label{fig:meron_shapes}
\end{figure}

{\em Merons: } So far we discussed magnets with an easy-axis anisotropy.  For an easy-plane anisotropy ($K < 0$), the topological defect with $Q = 1$ is a bound state of  vortex and antivortex (see Fig.~\ref{fig:meron_shapes}).  The sign of the out-of-plane magnetization in the core of the vortex is opposite to that in the antivortex core, so that each half of the skyrmion, called meron, has  $Q = \frac{1}{2}$.

The emergence of vortices and antivortices is related to the spontaneous breaking of $O(2)$ rotational symmetry by the uniform  in-plane magnetization. The bi-meron configuration in the FM state with $S_x = + 1$ can to a good approximation be obtained from the skyrmion configuration for the $S_z = +1$ state by  $\pi/2$-rotation around the $y$ axis: $(S_x, S_y, S_z)\rightarrow (S_z, S_y, -S_x)$ (see  Supplementary material for details), which explains the similarity between the stability diagrams for skyrmions and meron pairs in the $J_2-J_3$ and $J_2-K$ planes (Figs. \ref{fig:phase_diagr}(a,b) and Fig. 2 of Supplemental material).

\begin{figure}
\includegraphics[scale=0.32]{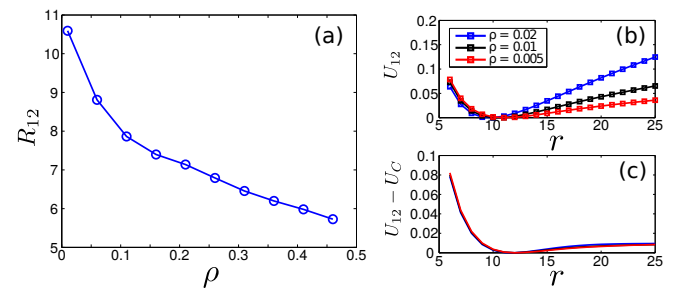}
\caption{(a) Optimal distance between the centers of merons,  $R_{12}$,  versus spin stiffness. $R_{12}$ increases near the LP,  $\rho=0$, resulting in a  ``fractionalization'' of topological charge.
(b) The meron-meron interaction potential, $U_{12}$, versus the distance between merons. (c) The ``non-Coulomb'' part of the interaction potential, $U_{12} - U_C$, versus meron-meron distance. 
 The model parameters are: $K=-5\cdot 10^{-3}$ are  $J_3=0.1$. The minimum of the potential energy curves is shifted to zero.
}\label{fig:R_2mer}
\end{figure}

Near a  LP the distance between merons is large  giving rise to two distinct peaks in the distribution of topological charge density  (Figs. \ref{fig:meron_shapes} (a,d,c,f)). The fractionalization of  skyrmion occurs because the two-dimensional Coulomb potential 
that confines vortex to an antivortex~\cite{Kosterlitz1973},
\begin{equation}\label{eq:U_mer}
U_{C}(r) = 2\pi \rho \ln (r/r_0), 
\end{equation}
 $r_0$ being the meron radius, vanishes  at the LP, $\rho = 0$.

Figure \ref{fig:R_2mer}(a) shows that at zero temperature the deconfinement of merons is incomplete: at the LP the optimal distance between merons, $R_{12}$,  remains finite,  because the meron-meron interaction energy, $U_{12}(r)$, has a minimum even at $\rho = 0$ (Figs. \ref{fig:R_2mer}(b,c)). The bi-meron molecules will, however, dissociate at $T \cancel{=} 0$.

At large distances  bi-merons interact via two-dimensional dipole-dipole interactions~\cite{Gross1978} resulting in formation of multimeron bound states. For $Q \gg 1$, the minimal-energy  configuration is the square lattice formed by merons (Figs. \ref{fig:multi-Q} (c,d)), analogous to the simple cubic lattice of half-skyrmions in nuclear physics~\cite{Kugler1989}.


{\em Conclusions: } We showed that metastable skyrmions and merons can exist in two-dimensional ferromagnets with conventional centrosymmetric lattices. Magnetic frustration required for stabilization of these topological excitations is considerably weaker than the one that destabilizes the FM state. Skyrmions in easy-axis magnets attract each other and in absence of magnetic field form long lines or rings facilitating their observation. In easy-plane magnets, skyrmions transform into pairs of merons, which dissociate at the Lifshitz point. Our results are directly relevant for square-lattice ferromagnets~\cite{Shannon2004}, and, qualitatively, they hold for other  lattice types including layered antiferromagnets with a weak AFM coupling between FM layers. In particular, bi-merons can exist in the collinear phase of the easy-plane triangular antiferromagnet NiBr$_2$~\cite{Day1976}. 
The bound states of magnetic skyrmions and merons show similarities to the skyrmions with large baryon numbers and crystals of half skyrmions in the original Skyrme model.
  
The authors are grateful to L. Balents for fruitful discussions. MM acknowledges the hospitality of UNSW and the FOM financial support.

\end{document}